\documentclass[journal=jctcce,manuscript=article]{achemso}

\usepackage[version=3]{mhchem}

\author{Carlos M. Bustamante}
\email{carlos.bustamante@mpsd.mpg.de}
\author{Franco P. Bonaf\'{e}}
\affiliation{Max Planck Institute for the Structure and Dynamics of Matter and Center for Free-Electron Laser Science, Luruper Chaussee 149, Hamburg 22761, Germany}
\author{Maxim Sukharev}
\affiliation{College of Integrative Sciences and Arts, Arizona State University, Mesa, Arizona 85212, United States}
\alsoaffiliation{Department of Physics, Arizona State University, Tempe, Arizona 85287, United States}
\author{Michael Ruggenthaler}
\affiliation{Max Planck Institute for the Structure and Dynamics of Matter and Center for Free-Electron Laser Science, Luruper Chaussee 149, Hamburg 22761, Germany}
\author{Abraham Nitzan}
\affiliation{Department of Chemistry, University of Pennsylvania, Philadelphia, Pennsylvania 19104, United States}
\author{Angel Rubio}
\affiliation{Max Planck Institute for the Structure and Dynamics of Matter and Center for Free-Electron Laser Science, Luruper Chaussee 149, Hamburg 22761, Germany}

\title[]
  {Molecular polariton dynamics in realistic cavities}

\begin{document}




\begin{abstract}
The large number of degrees of freedom involved in polaritonic chemistry processes considerably restricts the systems that can be described by any \textit{ab initio} approach, due to the resulting high computational cost. Semiclassical methods that treat light classically offer a promising route for overcoming these limitations. In this work, we present a new implementation that combines the numerical propagation of Maxwell’s equations to simulate realistic cavities with quantum electron dynamics at the density functional tight-binding (DFTB) theory level. This implementation allows for the simulation of a large number of molecules described at the atomistic level, interacting with cavity modes obtained by numerically solving Maxwell’s equations. By mimicking experimental setups,
our approach enables the calculation of transmission spectra, in which we observe the corresponding polaritonic signals. In addition, we have access to local information, revealing complex responses of individual molecules that depend on the number, geometry, position, and orientation of the molecules inside the cavity.
\end{abstract}

\section{Introduction}
Over the past decade, we have observed a rapid growth in the field of polaritonic chemistry \cite{ebbesen2023introduction}, focused on the strong coupling of molecular systems with the photonic modes of optical cavities. This interactions generates a hybrid state of light and matter known as polariton. For such strong light-matter coupling setups, it has been observed that it is possible to quench or accelerate chemical reactions, change their equilibrium constant, or modify their products \cite{garcia2021manipulating, nagarajan2021chemistry, xiang2024molecular}. To explain these phenomena in depth, theoretical polaritonic chemistry has made significant progress in recent years, which has been summarized in several reviews \cite{fregoni2022theoretical, ruggenthaler2023understanding, mandal2023theoretical}.

A comprehensive description of events occurring inside a cavity must account for many interacting components. Reactants giving rise to a reaction can interact with other reactants or solvent molecules through the photon modes within the cavity. Incorporating all these elements into a single theoretical framework with an appropriate description is a demanding task. The most accurate methodologies originate from \textit{ab initio} quantum electrodynamics (QED) approaches such as QED density functional theory (QEDFT)\cite{ruggenthaler2014quantum} and 
QED coupled cluster theory\cite{haugland2020coupled}. These methods are based on the non-relativistic Pauli-Fierz (PF) Hamiltonian where light and matter degrees of freedom (and the interaction between them) are treated quantum mechanically. However, they are typically restricted to a single photon mode, small molecules, and the use of the electric dipole approximation due to the high computational cost involved. Cavity Born-Oppenheimer dynamics is another \textit{ab initio} methodology based on the PF Hamiltonian \cite{flick2017cavity, schnappinger2023cavity, sidler2024unraveling}. In this approach, the displacement coordinate of the cavity modes is treated as a classical parameter in the same way as the position of the nuclei are treated in conventional Born-Oppenheimer calculations. Although this approach significantly reduces computational costs, enabling the description of larger systems relevant to chemistry, it is often used along with a few modes and the electric dipole approximation.

An alternative approach can be derived directly from the PF Hamiltonian in the semiclassical limit where the light degrees of freedom are considered classical and solved by Maxwell's equations, while representing the matter system quantum mechanically. In the same way, QEDFT can be taken to this limit, and after ignoring photon fluctuation functionals, the Maxwell-time-dependent density functional theory (Maxwell-TDDFT) is obtained \cite{jestadt2019light}.

With a much lower computational cost, semiclassical methods have proven to be highly convenient for working with various types of light fields \cite{bonafe2025full, albar2025high} and for incorporating macroscopic optical devices into simulations \cite{sukharev2023dissociation, sukharev2023efficient, sukharev2025unveiling}. Despite the limitations inherent to any mean-field method, semiclassical simulations have successfully reproduced quantum phenomena such as spontaneous and collective emission \cite{chen2019ehrenfest+1, bustamante2021dissipative, bustamante2022tailoring}, black-body thermalization \cite{gadea2022radiative}, vacuum effects \cite{hoffmann2019capturing, hoffmann2019benchmarking}, and others \cite{chen2019ehrenfest+2, tarasi2024interplay}; showing the great potential of this approach. Among these methods, the multi-trajectory Ehrenfest approach, developed by Hoffmann \textit{et al.}, is noteworthy \cite{hoffmann2019capturing, hoffmann2019benchmarking, hoffmann2020effect, rosenzweig2022analysis}. This enables the recovery of photon quantum information through post-processing of multiple semiclassical trajectories, where classical photon modes are sampled from the Wigner distribution. A potential step forward for this methodology would be to incorporate realistic optical environments and molecular systems represented at an atomistic level, which is relevant for polaritonic chemistry.

In the context of semiclassical methods, here we introduce a new implementation that combines the numerical propagation of Maxwell's equations to simulate realistic cavities, following the mean field approach of Sukharev \textit{et al.} \cite{sukharev2011numerical, sukharev2025unveiling}, with quantum molecular dynamics at the density functional tight-binding (DFTB) theory level \cite{elstner1998self}. This approach shares similarities with previous works where Maxwell's equations are coupled to Schr\"{o}dinger equation, described by tight-binding Hamiltonians \cite{lopata2009multiscale, li2025fdtd, trivedi2017model, smith2017capturing}. Nevertheless, the use of DFTB allows the atomistic representation of arbitrary molecules, making this tool useful to treat a wide variety of systems. Furthermore, since DFTB is an approximation of density functional theory, our approach follows the same theoretical framework of Maxwell-TDDFT \cite{jestadt2019light}.

We have simulated a Fabry-P\'{e}rot cavity by propagating Maxwell's equations on a grid with two parallel metal-like mirrors of specified thickness and dielectric response. Within the cavity, the electronic density of the molecules interacting with the Maxwell fields is time-propagated using real-time time-dependent DFTB \cite{Niehaus2009a,bonafe2020real}. In this work we focus on externally driven systems, and we analyze the evolution of individual and collective molecular properties, providing insights into the dynamics of the coupled systems. Moreover, we show how our implementation enables the study of collective and local effects and how these are affected by the position and orientation of the molecules. In this work, we restrict the description of Maxwell's fields to one (1D) and two dimensions (2D), while simulating the electron dynamics of nitrogen molecules (N$_2$), whose geometry and orientation are straightforward to control. Importantly, the molecule is always represented in three dimensions, despite the reduced dimensionality of the Maxwell fields. The results presented here aim to be the initial steps toward a more realistic representation of the cavity-molecule setup, ensuring reliability and maintaining a reasonable computational cost.

In Section \ref{sec:method}, we provide the details of our implementation. Sections \ref{sec:spatial1d} and \ref{sec:orient1d} focus on the study of collective effects and the spatial and orientational dependence of the local effects under electronic strong coupling, using the 1D setup. In Section \ref{sec:offres1d} we analyze the presence of heterogeneities in the molecular ensemble. Finally, in Section \ref{sec:spatial2d}  we investigate the more involved 2D case in a spatially resolved manner.
We conclude with our findings and future outlook in Section \ref{sec:conclusions}.

\section{\label{sec:method} Method and computational implementation}

As mentioned above, our approach can be understood as an approximation of QEDFT in the semiclassical limit \cite{ruggenthaler2014quantum, jestadt2019light}. Because of this, all the equations of motion presented here can be derived directly from PF Hamiltonian \cite{jestadt2019light} (for further details see Section S1 of the SI).
\begin{figure}
\includegraphics[scale = 0.5]{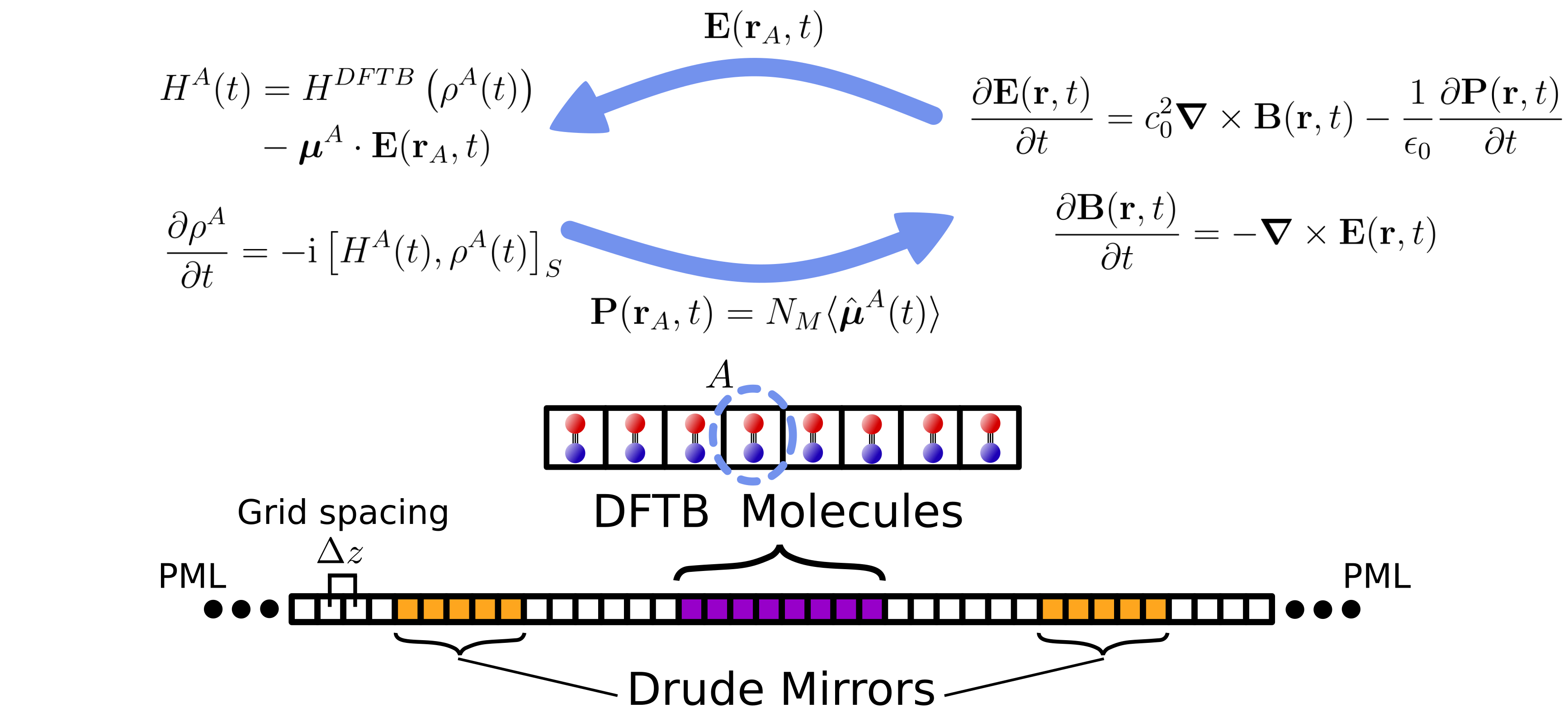}
\caption{\label{fig:scheme1} The upper part illustrates the forward–backward interaction between the propagation of Maxwell’s equations and the electronic density of a molecule labeled as $A$. The latter evolves according to the Liouville-von Neumann equation for a non-orthogonal basis (indicated by the subscript $S$). The bottom part shows the setup for the Maxwell propagation restricted to one (macroscopic) dimension, with open boundary conditions described by the perfectly matched layer (PML) method.}
\end{figure}
In our implementation, light and matter are treated in different scales and in a self-consistent way (Figure \ref{fig:scheme1}). The electromagnetic field is propagated numerically on a grid at the macroscopic scale by solving Maxwell's equations,
\begin{equation}
\label{eq:mxll_full}
    \begin{aligned}
        \frac{\partial \mathbf{B}(\mathbf{r}, t)}{\partial t} &= - \boldsymbol{\nabla} \times \mathbf{E}(\mathbf{r}, t), \\
        \frac{\partial \mathbf{E}(\mathbf{r}, t)}{\partial t} &= c_0^2 \boldsymbol{\nabla} \times \mathbf{B}(\mathbf{r}, t) - \frac{1}{\epsilon_0} \frac{\partial \mathbf{P}(\mathbf{r}, t)}{\partial t}.
    \end{aligned}
\end{equation}
Here, the macroscopic polarization
\begin{equation}
    \mathbf{P} = \bf{P}_{\rm{mir}} + \bf{P}_{\rm{mol}},
\end{equation}
is given by the response of either the classical mirrors ($\bf{P}_{\rm{mir}}$) or the molecules ($\bf{P}_{\rm{mol}}$), placed in particular positions on the grid. The response of the mirrors is simulated using the Drude model,
\begin{equation}
\label{eq:drude}
    \frac{\partial^2 \mathbf{P}_{\rm{mir}}(\mathbf{r}, t)}{\partial t^2} + \gamma \frac{\partial \mathbf{P}_{\rm{mir}}(\mathbf{r}, t)}{\partial t} = \varepsilon_0 \Omega_p^2 \mathbf{E}(\mathbf{r}, t),
\end{equation}
where $\Omega_p$ and $\gamma$ are the plasma frequency and the damping term, respectively. 

At the microscopic scale, the electron dynamics for each molecule is simulated by real-time time-dependent density functional tight-binding (TD-DFTB) \cite{elstner1998self, Niehaus2009a}, as implemented in the DFTB+ package \cite{bonafe2020real, hourahine2020dftb+}. The DFTB formalism is a self-consistent tight-binding model based on density functional theory \cite{elstner1998self}. Here we have used the implementation of real-time TD-DFTB in DFTB+ as an external library to propagate the electronic density of our molecules while coupling them to the Maxwell solver. We have chosen this method due to its good compromise between accuracy and computational cost. 

For our calculations we used the mio-1-1 Slater-Koster parameters \cite{elstner1998self}. In what follows, the indices $I,J$ correspond to the atoms on which the orbitals $i,j$ are centered, respectively, and the Greek indices $\alpha,\beta$ represent molecular orbitals. We do the the real-time propagation of the density matrix $\rho$ for each molecule $A$, given by the Liouville-von Neumann equation for a non-orthogonal basis \cite{bonafe2020real}:
\begin{equation}
    \label{eq:dftbrho}
    \dot{\rho}^A=-\mathrm{i}\left(S^{-1} H^A[\rho](t) \rho^A-\rho^A H^A[\rho](t) S^{-1}\right).
\end{equation}

The overlap and the time-dependent self-consistent Hamiltonian matrices are defined by $S_{ij}=\langle \phi_i | \phi_j \rangle$, $H_{ij}(t)=\langle \phi_i | \hat{H}(t) | \phi_j \rangle$. The initial density matrix $\rho$ is calculated from the ground-state eigenvectors $\{C_{\alpha i}\}$ and the molecular orbital occupations $f_\mu$: $\rho_{ij}(0)=\sum_\alpha C_{\alpha i} f_\alpha C^*_{\alpha j}$. The time-dependent self-consistent DFTB Hamiltonian matrix $H[\rho](t)$ is calculated as: 
\begin{equation}
\begin{aligned}
    \label{eq:dftbham}
    H_{i j}^A[\rho](t) &= H^{0,A}_{i j}+\frac{1}{2} S_{i j}\sum_K (\gamma_{IK} + \gamma_{JK})\Delta q_K(t) \\
    &- \frac{1}{2} S_{i j}(\boldsymbol{\mu}_{I} + \boldsymbol{\mu}_{J}) \cdot \mathbf{E}(\mathbf{r}_A,t)).
\end{aligned}
\end{equation}
In this expression $H^{0,A}$ is the non-self-consistent Hamiltonian for molecule $A$, $\gamma$ is a distance-dependent function used for the Coulomb interaction term and $\boldsymbol{\mu}$ is the dipole moment matrix. As it can be seen, the time-dependence of the Hamiltonian comes both from the Coulomb interaction term (second term in the right-hand side of Eq. \ref{eq:dftbham}), as well as by the external potential due to the dipole coupling to the local electric field at the position of the molecule $A$ (third term on the right-hand side of Eq. \ref{eq:dftbham}). More details on the DFTB Hamiltonian can be found in references \cite{elstner1998self, Frauenheim2002b, Niehaus2009a}. The Coulomb interaction in turn depends on the time-dependent density via the net Mulliken charges $\{\Delta q_I\}$, defined as $\Delta q_I(t) = q_I(t) - q_I^0$, where $q_I^0$ is the effective gross charge of the neutral atom and $q_I(t)$ is calculated as
\begin{equation}
    \label{eq:dftbq}
    q_I(t)=\sum_{j \in I} \sum_i \rho_{i j}(t) S_{i j}.
\end{equation}

With this, the molecular component of the polarization coming from the molecule $A$ and following the mean field approximation, can be calculated as
\begin{equation}
\label{eq:polarization}
    \mathbf{P}_{\rm{mol}}(\mathbf{r}_A) = N_M\, \langle \hat{\boldsymbol{\mu}}^A \rangle,
\end{equation}
where $\langle \hat{\boldsymbol{\mu}}^A \rangle$ is the expectation value of the dipole moment of the molecule $A$ at its corresponding grid point located at $\mathbf{r}_A$, and $N_M$ is the molecular number density \cite{sukharev2023dissociation, sukharev2023efficient, sukharev2025unveiling}. For a continuous spatial distribution of molecules, $N_M$ must satisfy the inequality
$N_M<1/\Delta x^3$, where
$\Delta x$ is the grid spacing. This condition ensures the consistency of physical properties across neighboring spatial positions. In the case where the molecules are treated as Dirac-delta emitters, we must set
$N_M=1/\Delta x^3$ \cite{sukharev2025unveiling}. In all cases, we assume that the molecules are located at the centers of their respective grid points, and that the distances between molecules are large enough to justify the approximation of intermolecular interactions using the mean field approach. These interactions are always time-retarded due to the spatial distribution of the molecules on the grid.

When the propagation of the electromagnetic fields is done in one dimension, the general three-dimensional equations reduce to
\begin{equation}
    \label{eq:mxll_1d}
    \begin{aligned}
        \frac{\partial B_y(z,t)}{\partial t} &= - \frac{\partial E_x(z,t)}{\partial z}, \\
        \frac{\partial E_x(z,t)}{\partial t} = -c_0^2 &\frac{\partial B_y(z,t)}{\partial z} -\frac{1}{\epsilon_0} \frac{\partial P_x(z,t)}{\partial t},
    \end{aligned}
\end{equation}
considering that the fields propagate in the $z$-direction. If we restrict the propagation of the fields to two dimensions we can work with the set of equations known as \textit{the transverse-magnetic mode with respect to z} (TM$_z$) \cite{taflove2005computational},
\begin{equation}
\label{eq:TMz}
    \begin{aligned}
        &\frac{\partial B_x(\mathbf{r},t)}{\partial t} = - \frac{\partial E_z(\mathbf{r},t)}{\partial y}, \\
        &\frac{\partial B_y(\mathbf{r},t)}{\partial t} = - \frac{\partial E_z(\mathbf{r},t)}{\partial x}, \\
        \frac{\partial E_z(\mathbf{r},t)}{\partial t} = &c_0^2 \left( \frac{\partial B_y(\mathbf{r},t)}{\partial x} - \frac{\partial B_x(\mathbf{r},t)}{\partial y}\right) -\frac{1}{\epsilon_0} \frac{\partial P_z(\mathbf{r},t)}{\partial t},
    \end{aligned}
\end{equation}

Equations \ref{eq:mxll_1d} and \ref{eq:TMz} are propagated numerically using the finite difference time domain (FDTD), method \cite{taflove2005computational}, using the implementation of Sukharev \textit{et al.} \cite{sukharev2023dissociation, sukharev2023efficient, sukharev2025unveiling}. As shown in the scheme for our 1D implementation in Figure \ref{fig:scheme1}, two sections of this grid are allocated for the mirrors, while the central region is reserved for the molecules. Even though we propagate Maxwell's equations in 1D and 2D, the molecules are always represented in three dimensions.

We also implemented open boundary conditions for the propagation of Maxwell equations by using the convolutional perfectly matched layer (CPML) method \cite{roden2000convolution}. By doing so we can work with all the free-space modes that can be captured by our grid resolution.

\section{\label{sec:results} Results}

\subsection{Spatial dependence in a 1D cavity}
\label{sec:spatial1d}

In classical optics, the geometry of the cavity/photonic device determines the frequency and the shape of the electromagnetic modes. By controlling the geometry, it is possible to tune the frequency of the cavity modes to resonate with an electronic transition (or another kind of transition) of a molecular ensemble. The simplest example is the Fabry-P\'{e}rot (FP) cavity, conformed by two parallel mirrors. Varying the distance between them one can control the frequency of the cavity eigenmodes. This setup is mimicked by our 1D implementation. Separating the mirrors by 123 nm, we get a FP cavity whose third mode is in the order of the electronic transition of the N$_2$ molecule, predicted by DFTB. At this level of theory, we calculated an equilibrium bond length of 110.77 pm and a first electronic transition energy of 13.902 eV, corresponding to a wavelength of 89.18 nm. For this example we placed all the molecules in parallel respect to the the x-component of the electric field ($E_x$). The microscopic three-dimensional simulation boxes for the individual molecules are represented by their respective positions and dipole moments in the macroscopic Maxwell grid. Thus, for a fixed $N_M$, the more molecules we put into the macroscopic grid, the more space the molecules will occupy in this grid. We fill the cavity volume symmetrically from the middle (see Figure \ref{fig:scheme1} for a graphical illustration).
The mirrors are modeled using the Drude model (Equation \ref{eq:drude}) for which we choose the parameters $\Omega_p=$ 34 eV and $\gamma=$ 0.181 eV, and the width of the mirrors is 20 nm. These values do not represent a real material, but are merely chosen for presentational convenience, as they provide reasonable reflectivity at the working frequencies, enabling the system to reach the strong-coupling regime. All the results presented in this subsection are obtained with a grid spacing $\Delta z =$ 1 nm, integration time step of $\Delta t_{\mathrm{Mxll}}=$ 2.419$\times 10^{-4}$ fs for the Maxwell system and $\Delta t_{\mathrm{mol}}= 2\Delta t_{\mathrm{Mxll}}$ for the propagation of the electronic density, and total propagation time of 1000 fs. The molecular number density (Equation \ref{eq:polarization}) is set to $N_M$ = 6.75$\times 10^{-2}\,\mathrm{nm}^{-3}$, which is equivalent to a molar concentration of 0.112 M.

The system is excited by an external pulse that comes from the right side of our simulation box with a frequency $\omega_l$ such that $\hbar \omega_l = 14$ eV, and a pulse duration of 2 fs. The spectral broadening of this pulse mainly excites the polaritons, with only a minor response from the closest non‑resonant cavity modes.
We collect the light transmitted in some point between the left mirror and the the CPML section, $z=z_{\rm det}$, being $z_{\rm det}$ the position where the signal is detected. The transmission spectrum is calculated according to \cite{taflove2005computational}
\begin{equation}
    P(z_{\rm det}, \omega) = - \mathrm{Re}\{ E_x^* (z_{\rm det}, \omega) H_y(z_{\rm det}, \omega) \}  
\end{equation}
\begin{figure}[t]
    \centering
    \includegraphics[scale=0.26]{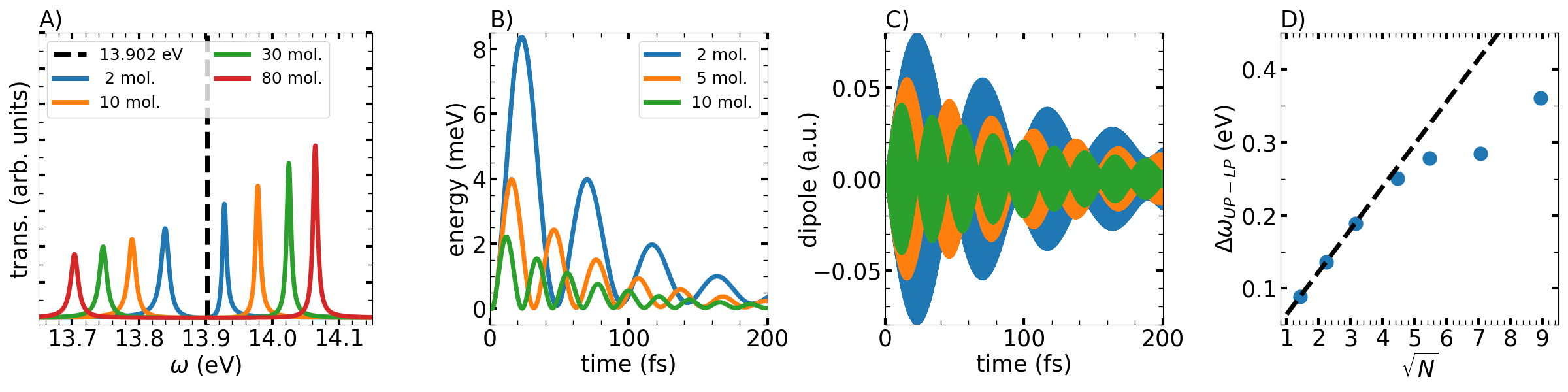}
    \caption{A) Transmission spectra obtained after exciting the cavity and different number of simulated molecules. The black dashed line indicates the first electronic excitation energy of N$_2$. Every molecule is placed in individual grid points, next to the other, separated by 1 nm. B) Energy evolution and C) dipole evolution of the N$_2$ molecule placed in the center of the cavity with different number of simulated molecules. D) Rabi splitting as a function of the square root of the number of simulated molecules.}
\label{fig:ener_dip_spec_rabi}
\end{figure}
In Figure \ref{fig:ener_dip_spec_rabi}A we observe the result spectra for different numbers of simulated molecules. The Rabi splitting indicates that we reach the strong coupling regime. As the number of molecules $N$ increases, their coherent dipole oscillations enhance the coupling strength, increasing the Rabi splitting. The asymmetry in the intensities shows a higher population of the upper polariton (UP) compared to the lower polariton (LP). This is a consequence of the spectrum of the incoming laser as well as the frequency-dependent response of the mirrors \cite{sukharev2023dissociation}.

Our implementation allows us to track the evolution of the energy and dipole moment of each simulated molecule inside the cavity.
For example, in Figure \ref{fig:ener_dip_spec_rabi}B and C we show these observables for the molecule at the center of the cavity, considering three different conditions that differ in the total number of molecules. Here, the strong coupling is reflected in the energy oscillations (Rabi oscillations) and dipole moment modulation. The larger the number of molecules, the higher the frequency of the energy oscillations. The maximum absorbed energy also decreases as additional molecules also absorb light and the collective coupling accelerates the emission rate.

Typically, the Rabi splitting scales linearly with the square root of the number of molecules, assuming all molecules experience the same electric field. However, in our case, each molecule is subject to different field amplitudes depending on its position. Figure \ref{fig:ener_dip_spec_rabi}D shows that for a small number of molecules, the space occupied by them is less than half the wavelength of the corresponding excitation energy. Under these conditions, the electric field amplitudes at the position of each molecule is almost constant, resulting in a linear dependence of the Rabi splitting respect to $\sqrt N$. For a larger number of molecules (e.g. 30 or 50 for our simulation setup) the space occupied by the molecular ensemble becomes comparable to the wavelength of the resonant cavity mode. In this case, the usually assumed $\sqrt{N}$ behavior no longer holds, since not every molecule contributes in the same way to the Rabi splitting. We thus see that the spatial distribution can have a profound impact on these relation. Interestingly, due to the spatial form of the resonant cavity mode, if we increase even more the number of molecules (e.g. 80 molecules) we can again observe a regime with a $\sqrt{N}$ dependence. That is due to the fact that the resonant mode is the third mode of the cavity, hence it has two nodes (in this one-dimensional Fabry-Perot case), and the molecules located close to the other mode maxima start to contribute equally again. We can show this explicitly by studying the dipole response of each individual molecule.
\begin{figure}[t]
    \centering
    \includegraphics[scale=0.28]{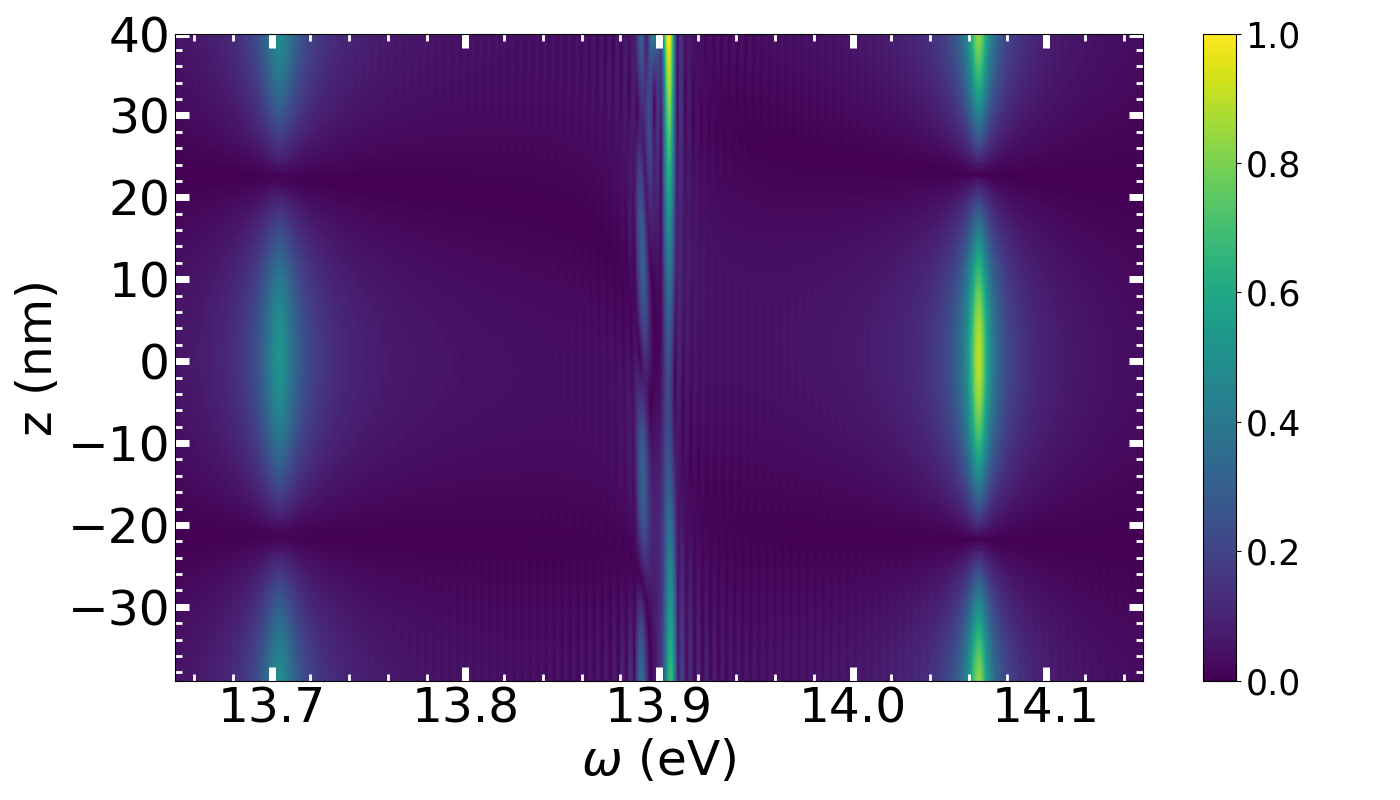}
    \caption{Fourier-transformed dipole moments of the simulated molecules as a function of their position in the cavity, for the system comprising 80 molecules. The color scale has been normalized with the maximum amplitude of the dipole moment.}
    \label{fig:dip_2D_1}
\end{figure}
In Figure \ref{fig:dip_2D_1}, we present the Fourier transform of the dipole moment for every simulated molecule when 80 molecules are placed in the middle of the cavity. The polaritonic peak intensities (at 13.7 eV and 14.06 eV) exhibit a spatial distribution that matches the expected profile of the cavity's third mode, with nodes near one-third and two-thirds of the mirror separation distance. We note that despite the separation distances among molecules, the cavity makes them oscillate coherently. This shows explicitly how the cavity introduces new length scales into the ensemble and lets otherwise uncorrelated molecules act in a correlated manner.

Experimental results have previously demonstrated the position-dependent nature of the Rabi splitting when a thin J-aggregate layer is displaced within the cavity \cite{wang2014quantum}. Depending on the position of the layer, the observed Rabi splitting changes following a mode-like profile. 
Beyond highlighting how the relative position of the maxima of the resonant mode and the molecular ensemble determines the Rabi splitting and the coherent response, our results show that all molecules present the same Rabi splitting, and they differ in the intensity of the polaritonic peaks, which depends on their position in the cavity and on the cavity mode shape. This is an important detail, which is not directly accessible experimentally. Yet it is very relevant for understanding how collective strong coupling can induce local effects. Since the first demonstration that collective strong coupling can induce local effects \cite{sidler2020polaritonic}, the search for such collective-to-local quantities has taken center stage in polaritonic chemistry.

Figure \ref{fig:dip_2D_1} also reveals two interesting features near $\omega = 13.9$ eV. The lower-frequency signal displays a mode-like structure with nodes at one-fourth, one-half, and three-fourths of the mirror separation distance. These oscillations arise from off-resonant coupling between the molecules and the fourth cavity mode, which also contributes to the observed redshift. This mode becomes weakly active due to the spectral width of the excitation pulse.
The intense line above 13.9 eV corresponds to the N$_2$ electronic excitation. While both features are found in the dipole spectra, they are negligible in the transmission spectrum, representing dark states of the system. In the semiclassical picture, these states emerge from molecular dipole oscillations that generate destructive interference patterns \cite{bustamante2022tailoring}, suppressing emission at the corresponding frequency. The variation of intensity of the dark state above 13.9 eV arises due to the asymmetrical excitation of the system. The molecules located closer to the mirrors are the first to absorb either the incoming pulse or reflections from the mirrors, and therefore they exhibit the strongest dark state peaks.

\subsection{Orientational effects in a 1D cavity}
\label{sec:orient1d}

In the first example, we have aligned all the molecules in order to analyze only the positional dependence. Now we consider a little more complex setup and allow for random molecular orientations, characterized by $\mathrm{cos}^2\theta$, where $\theta$ is the angle between the molecular bond and the polarization axis (x-axis). This allows to investigate the interplay between positional and orientational effects. For this part we reduced the grid spacing to $\Delta z= 0.1$ nm, maintaining the separation between molecules 1 nm. The 10 grid points of spacing between molecules guarantee convergence of the simulations \cite{sukharev2025unveiling}. We worked with $N_M= 6.75\times 10^{-1}\,\mathrm{nm}^{-3}$ wich ensures that the case of all the molecules placed in parallel to the x-axis reproduces the results presented in the previous section.
\begin{figure}[t]
    \centering
    \includegraphics[scale=0.65]{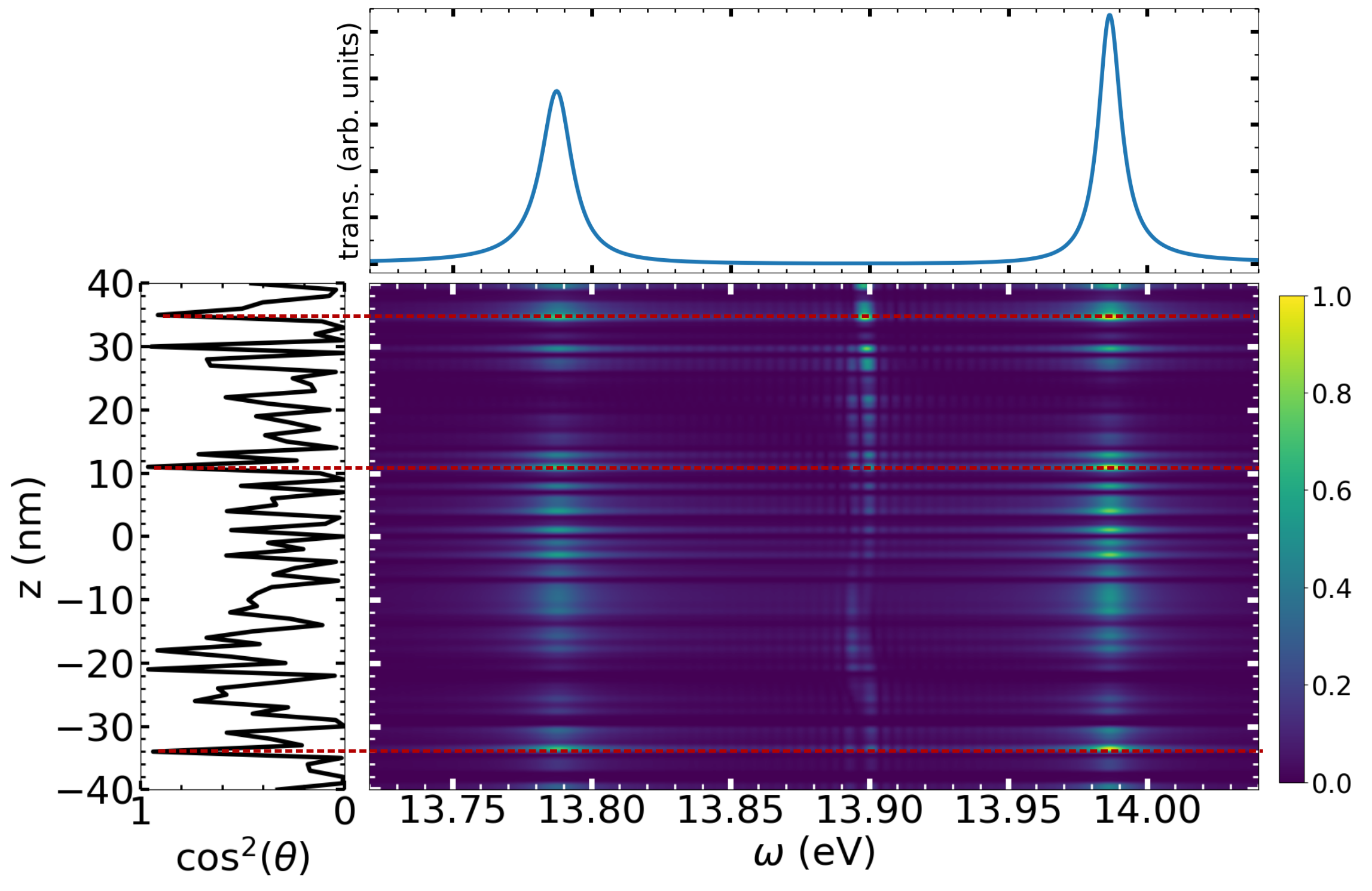}
    \caption{Density plot showing the Fourier transform intensity of the molecular dipole moment versus position in the cavity for 81 randomly oriented molecules The color scale has been normalized with the maximum amplitude of the dipole moment. Left panel: spatial distribution of cos$^2\theta$, where $\theta$ is the angle between the N$_2$-bond and x-axis. Upper panel: simulated transmission spectrum. Red dashed lines highlight molecular positions exhibiting the strongest polaritonic signals.}
    \label{fig:dip_2D_2}
\end{figure}
Figure \ref{fig:dip_2D_2} shows the results of our implementation for 81 molecules with random orientations. Only those with a non-zero dipole component along the $x$-axis can couple to the 1D setup of the cavity. The resulting transmission spectrum shows a Rabi splitting smaller than the expected for the same amount of ordered molecules. The Fourier transforms of individual dipole moments evidence the disorder. When present, polariton signals appear at the same frequencies as the transmission spectrum peaks. The red dashed lines in Figure \ref{fig:dip_2D_2} highlight molecules most aligned with the $x$-axis, which exhibit the strongest polaritonic intensities. Despite the disorder, molecules at nodal positions consistently lack polaritonic signals, regardless of their orientation. The peaks previously attributed to off-resonant interactions as well as dark-state formation are also affected in a similar manner by the orientation of the molecules. The results presented here are in qualitative agreement to those in \cite{fiechter2024understanding}. In 10 simulations with different molecular orientations, we obtained an average orientation $\langle \mathrm{cos}^2 \theta \rangle = 0.342$, close to the expected value of 0.33 for a large ensemble. The average Rabi splitting is 0.216 eV, which corresponds to a rescaling factor of 0.6, in contrast to the 0.7 ($1/\sqrt{2}$)  factor predicted by ref.~\cite{fiechter2024understanding}. This could be a consequence of the spatial dependence of the field-matter coupling in our description: since every molecule interacts differently with the cavity mode, the effect of disorder on the reduction of the Rabi splitting will be inhomogeneous, leading to a lower value than the one predicted for a space-independent coupling model of Ref.\cite{fiechter2024understanding}.

\subsection{Molecular heterogeneity in a 1D cavity}
\label{sec:offres1d}

Local effects arising from collective strong coupling can also be observed when introducing a heterogeneity in the molecular ensemble, resembling the presence of vibrational disorder. In our setup, inspired on the work of Sidler \textit{et al} \cite{sidler2020polaritonic}, this disorder takes the simple form of a single N$_2$ molecule with a different bond length of 111.12 pm (hereafter called "stretched" molecule), surrounded by numerous cavity-resonant N$_2$ molecules (called ``unstretched" ensemble). The calculated first electronic transition of the stretched molecule is 13.804 eV, which is off-resonant with any cavity mode. Again, we used a grid spacing of $\Delta z = $ 0.1 nm and a spacing between molecules of 10 grid points.

\begin{figure}[t]
    \centering
    \includegraphics[scale=0.3]{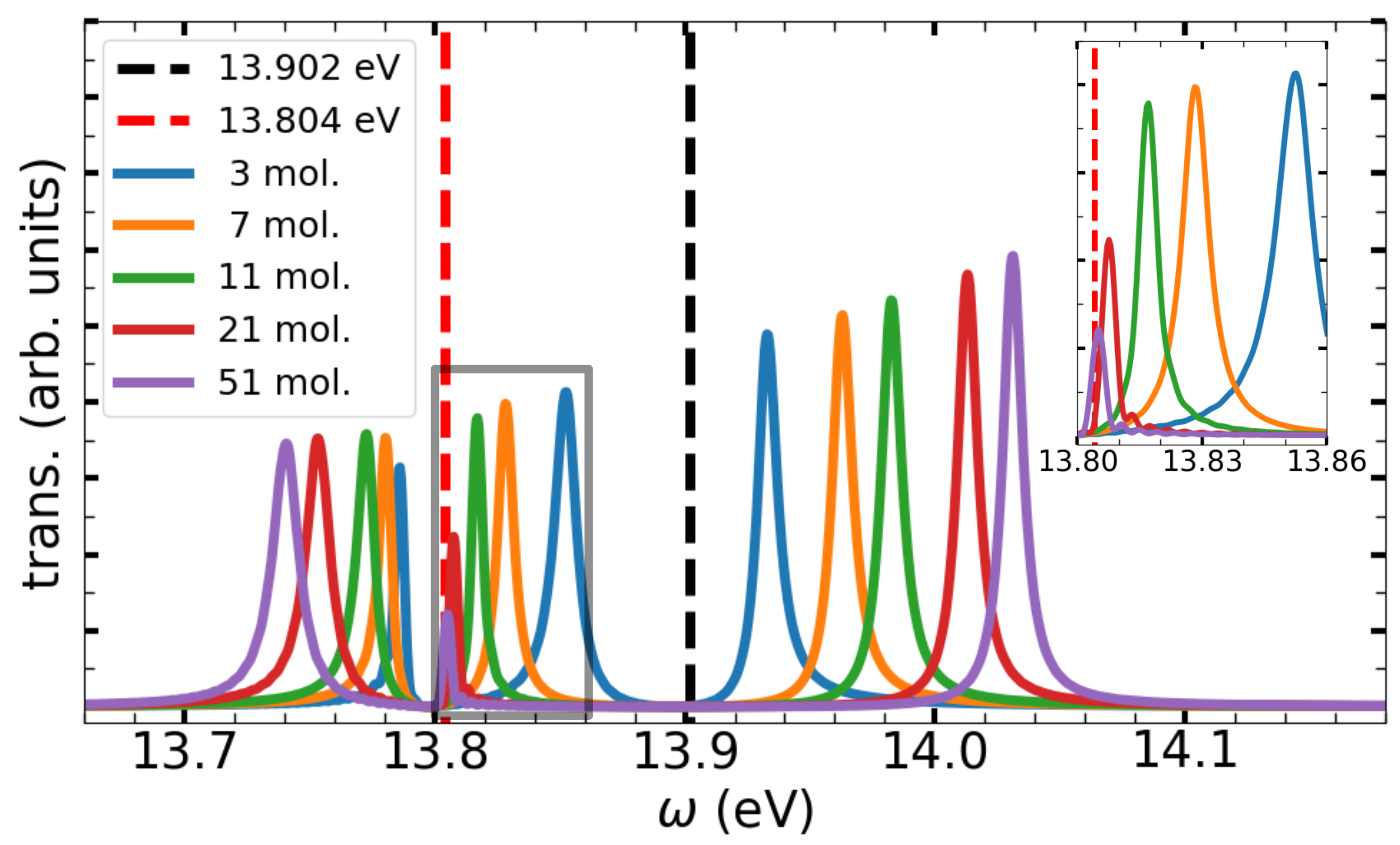}
    \caption{Transmission spectra obtained after exciting the cavity with varying numbers of simulated molecules, where one molecule has a different bond length. The black dashed line indicates the first electronic excitation energy of the unstretched N$_2$ (i.e. with the equilibrium interatomic distance), while the red line corresponds to the stretched molecule. The inset provides a zoomed-in view of the area indicated by the gray square.}
    \label{fig:spec3}
\end{figure}

The inclusion of this stretched molecule in the cavity generates an intermediate polaritonic signal in the spectra, as seen in Figure \ref{fig:spec3}. As the collective coupling strength increases, this signal becomes damped due to the dominant collective interactions, a phenomenon previously predicted by QED simulations \cite{sidler2020polaritonic}. 

\begin{figure}[t]
    \centering
    \includegraphics[scale=0.28]{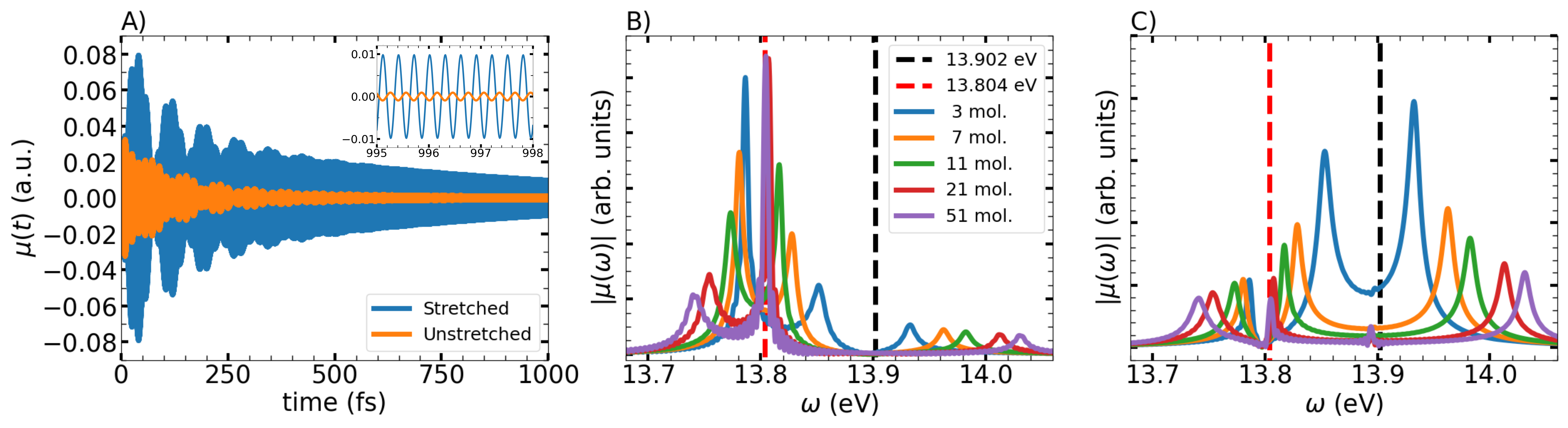}
    \caption{A) Evolution of the dipole moment for the stretched molecule (blue) compared to the average dipole moment of the unstretched molecules (orange) in a cavity containing 21 molecules. The inset zooms in the last stages of the simulation. B) Fourier transform of the stretched molecule's dipole moment within the cavity with different number of molecules. C) Same but for the first unstretched neighbor on the right.}
    \label{fig:dip_t_w}
\end{figure}

To further understand the spectral features, we compare in Figure \ref{fig:dip_t_w}A the dipole dynamics of the stretched molecule with the ensemble average of unstretched molecules following the cavity excitation. The unstretched molecule dipole rapidly decays and begins to oscillate out of phase with the stretched molecule, resulting in radiative quenching (see inset). At this stage, the stretched molecule presents an oscillation amplitude considerably larger than the ensemble average of the rest.

By Fourier analysis of these signals, the dipole spectra for the stretched (Figure \ref{fig:dip_t_w}B) and unstretched (Figure \ref{fig:dip_t_w}C) molecules are obtained. It is clear from panel (B) that the stretched molecule's contribution to the lower polariton diminishes with increasing molecular number, while its middle polariton signal strengthens. This panel also shows that the dipole of the stretched molecule remains highly active even when the middle polariton state gets darker, in a semiclassical analogy to the results of Ref.~\cite{sidler2020polaritonic}. Despite physical proximity, the immediate right neighbor (unstretched molecule) shows a very different behavior, as seen in panel (C). Namely, the signal associated with the middle polariton decreases with the number of molecules, becoming even lower than the signals of the upper and lower polaritons. Finally, comparison of Figure \ref{fig:dip_t_w}C with Figure \ref{fig:spec3} shows that, under strong collective coupling, the middle polariton signal has a predominant contribution from the stretched molecule, while the unstretched ensemble is responsible for the the upper/lower polariton signal. This resembles qualitatively the results obtained by full QED calculations \cite{sidler2020polaritonic}.

The setup also allows us to consider new scenarios, such as varying the number of stretched molecules, different stretching amplitudes, orientations, and positions. \textit{A priori}, increasing the number of stretched molecules should require a larger number of unstretched molecules to produce the same effect in the spectra. Cases in which the number of stretched molecules equals or exceeds the number of unstretched ones could lead to a blue-shift of the middle polariton. However, all these cases will also be influenced by the position and orientation of the molecules. These aspects will be addressed in future work.

\subsection{Spatial dependence in a 2D cavity}
\label{sec:spatial2d}

We now work with Maxwell equations in two-dimensions by the implementation of Equation \ref{eq:TMz}. The new setup is composed by a simulation box of 340 $\times$ 340 nm$^2$ placed in the xy-plane, where a 20 nm region on the boundaries is dedicated to the CPML. Again, we work with a FP cavity for which we set two parallel mirrors separated by a distance of 123 nm, as used in the previous sections. Each mirror has a width of 20 nm and a length of 280 nm.
We continue working with a grid spacing 
$\Delta x=\Delta y=$ 1 nm, and an integration time step of $\Delta t_{\mathrm{Mxll}}=$ 2.419$\times 10^{-4}$ fs for the Maxwell system and $\Delta t_{\mathrm{mol}}= 2\Delta t_{\mathrm{Mxll}}$ for the DFTB molecules.  The total integration time of our simulations was 500 fs.

\begin{figure}[t]
    \centering
    \includegraphics[scale=0.6]{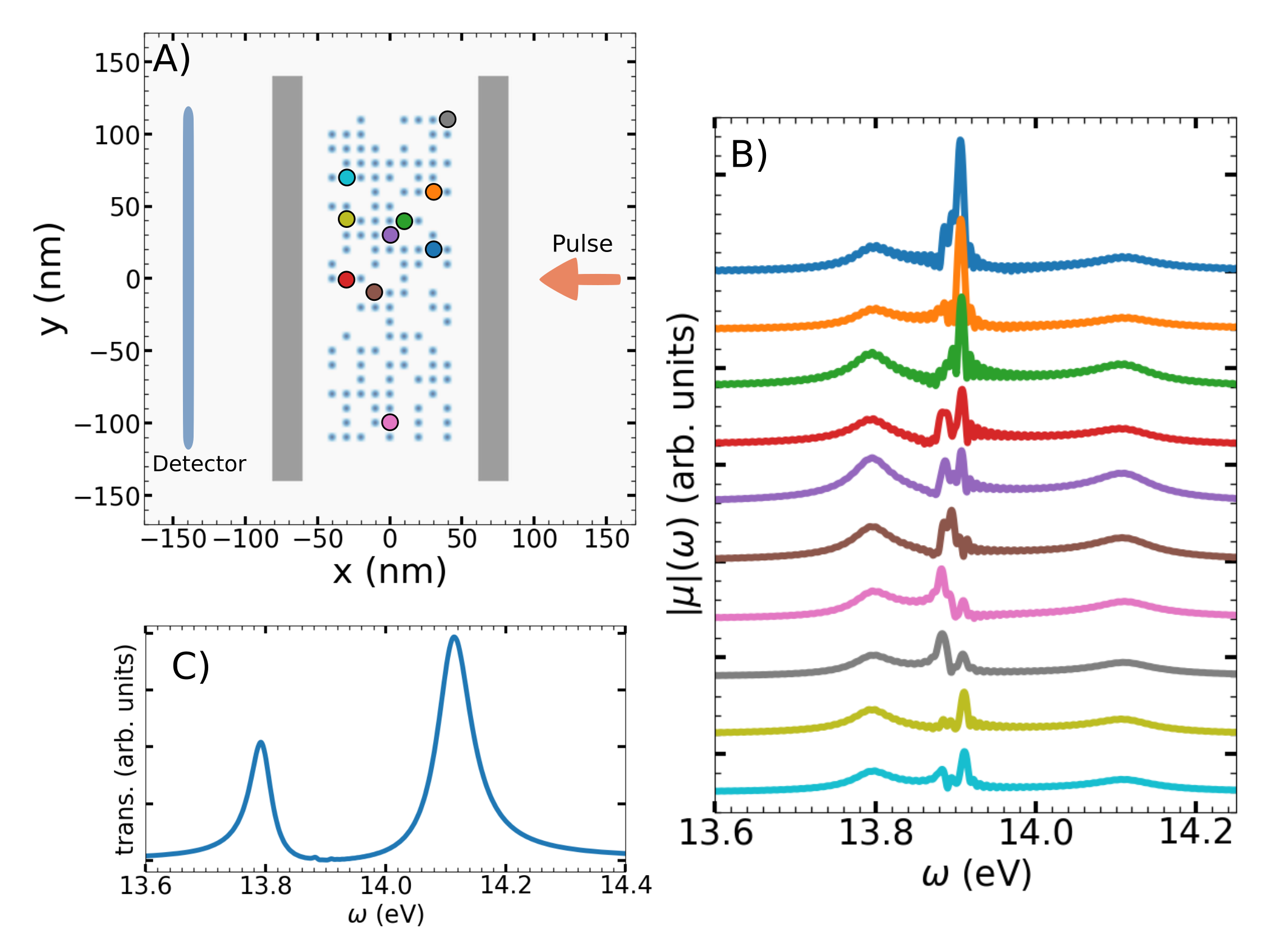}
    \caption{A) Schematic of the simulation box. Light-blue dots represent the position of the N$_2$ molecule, while gray rectangles denote mirror locations. B) Fourier transforms of the dipole moment of the highlighted molecules in panel A (colors correspond to the molecular position). C) Transmission spectra calculated from grid points along the detector line (shown in panel A).}
    \label{fig:2D_rab_dip}
\end{figure}

For the first analysis, we worked with 120 molecules oriented parallel to the z-axis. To maximize the space occupied by the molecules, we distributed them randomly inside a central area of 80 $\times$ 110 nm$^2$, as can be seen in Figure \ref{fig:2D_rab_dip}A. Their coordinate positions were set as $\mathbf{r}=(10\,a\;\mathrm{nm}, 10\,b\;\mathrm{nm})$, where $a$ and $b$ are random integer numbers. In order to avoid numerical issues, the macroscopic polarization (Equation \ref{eq:polarization}) calculated for each molecule was spatially smoothed by a normalized Gaussian function with a broadening equal to the grid spacing. 
Beyond 1D, the shape and the orientation of the electromagnetic source have a relevant effect on the type of mode that is activated inside the cavity. In order to mimic experimental conditions while avoiding scattering from the mirror edges we focus our source in the central part of the right mirror. For this
we used 9 point-like Gaussian pulses symmetrically placed on a line at $y = 120$ nm, symmetrically separated by 10 nm, with respect to $x = 0$ (Figure S1). Each pulse has a frequency of 14 eV and a Gaussian envelope with a FWHM of 0.588 fs.

The transmission spectra in the 2D setup  is calculated according to:
\begin{equation}
    P(x_{\rm det},\omega) = \mathrm{Re}\, \int_{y_{\rm min}}^{y_{\rm max}} E_z^*(x_{\rm det}, y, \omega) \, B_y(x_{\rm det}, y, \omega) dy,
\end{equation}
where the integral goes from $y_{\rm min}= -120$ nm and $y_{\rm max}= 120$ nm, corresponding to a line of grid points located on the left of the cavity, at $x=x_{\rm det} = -140 \, \rm nm$ (light-blue line in Figure \ref{fig:2D_rab_dip}A).

By Fourier-transforming every individual molecular dipole moment we can examine their spectral responses in detail. Figure \ref{fig:2D_rab_dip}B presents these spectra for selected molecules. The selection was arbitrary and aimed to cover various regions within the cavity. Notably, while the transmission spectrum (Figure \ref{fig:2D_rab_dip}C) displays only the two dominant polaritonic peaks, the molecular-level spectra show significantly richer and spatially dependent features. The individual spectra show signatures corresponding to both polaritonic states, the characteristic N$_2$ electronic transition (at 13.9 eV) and a redshifted signal due to off-resonant interactions with higher-order cavity modes (see section \ref{sec:spatial1d}).

\begin{figure}[t]
    \centering
    \includegraphics[scale=0.6]{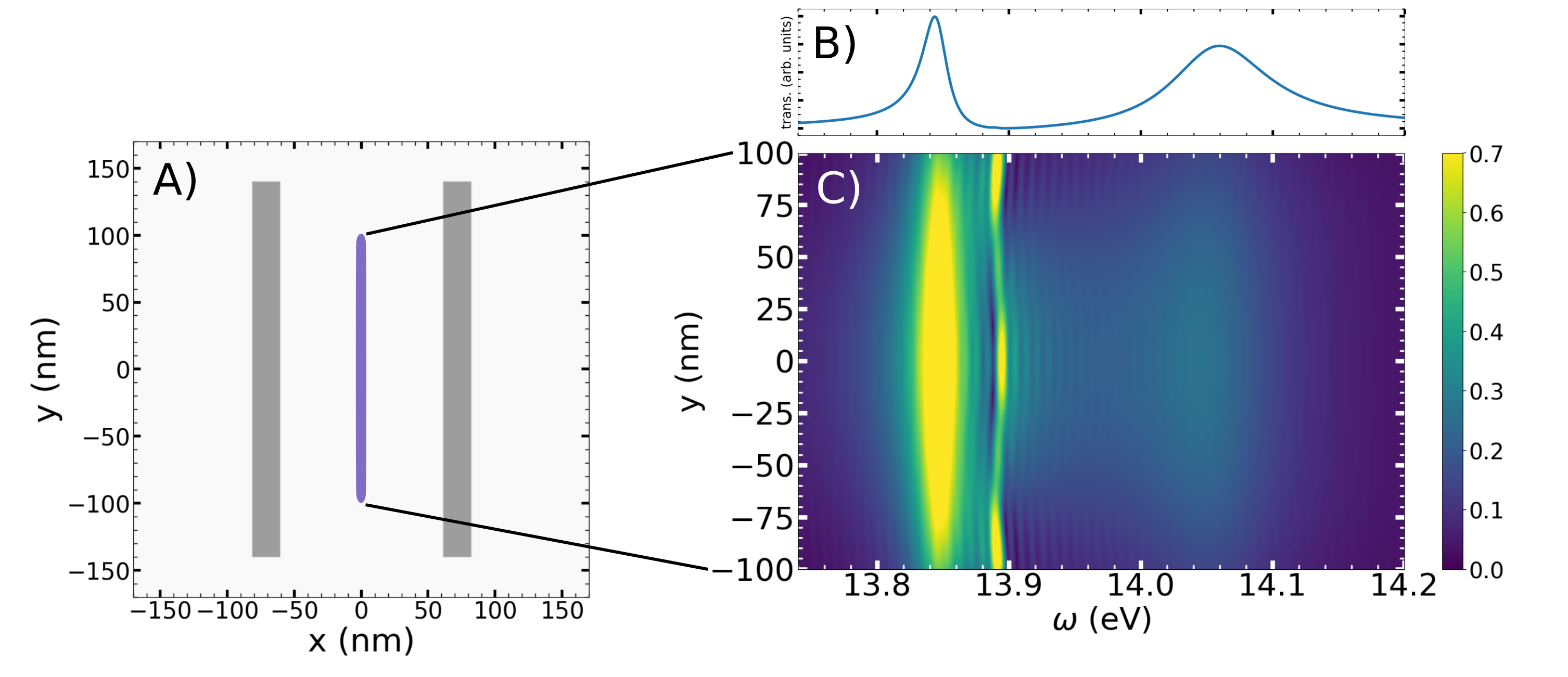}
    \caption{A) Diagram of our 2D cavity where the violet line indicates the position of the 201 nitrogen molecules placed along the x =0 line. B) Transmission spectrum obtained after exciting the cavity. C) The density map shows the amplitude of the Fourier transform of the dipole moment of every molecule according to the position}
    \label{fig:line_2D_dip}
\end{figure}

Similarly to the study of the spatially-resolved spectral response of the molecules in the 1D cavity, here we also study the response along the new vertical axis, where we the cavity present the highest losses. To this end, we performed a calculation in a modified setup with 201 molecules symmetrically aligned at the cavity center ($x = 0$), as shown in Figure \ref{fig:line_2D_dip}.
The transmission spectrum is depicted in Figure \ref{fig:line_2D_dip}B and again shows two polaritonic peaks. Spectral analysis of individual molecular dipole responses, shown in Figure \ref{fig:line_2D_dip}C, reveals that both upper (UP) and lower (LP) polaritons are present, but the UP has larger broadening and reduced intensity compared to the LP. Their intensities decay radially from the cavity center, reflecting the evanescent nature of the cavity mode (Figure S2). Unlike the previous case, here we observe only one dark state below 13.9 eV, which could be generated by an off-resonant interaction with another cavity mode, as we observed previously.

\section{Conclusions}
\label{sec:conclusions}

In this work, we have presented a new implementation that combines numerical propagation of Maxwell's equations using the FDTD method (in 1D and 2D) with molecular quantum dynamics at the DFTB level. This implementation enables the exploration of molecular polaritonic dynamics inside Fabry-P\'{e}rot cavities defined by two realistic mirrors, without any assumptions about the number of modes. As demonstrated throughout this work, our setup provides direct analysis of spectra from the electromagnetic fields, comparable to experimental observables, as well as the individual dynamics of each simulated molecule with spatial resolution. Additionally, we can understand the effects of orientation and geometry of the molecules simply by modifying the molecular geometry input files.

We have shown that the spectral information transmitted by the cavity does not give access to all the details of the molecular-level phenomena. Factors such as the number of molecules, their location, orientation, and geometry are all relevant to describe properly the processes occurring inside a realistic cavity.

We also highlight that the computational cost of these simulations is low; most of the calculations were completed within a few hours (or minutes) on a personal computer. Nevertheless, extending the approach to three dimensions or including larger molecular systems will require appropriate high performance computing methods, as demonstrated in ref.~\cite{sukharev2023efficient}.

Although not presented here, the use of DFTB+ as an external library also makes it possible to perform Ehrenfest and Born-Oppenheimer dynamics. In future works we will show the interplay of the different degrees of freedom with the cavity. Moreover, quantum photon effects can be included by the use of QEDFT functionals, which will be done in future implementations.

\section*{Data Availability}
Data will be made available upon request.

\section*{Supporting Information}
\begin{itemize}
    \item Additional details about the semi-classical limit of Pauli-Fierz Hamiltonian and QEDFT.
    \item Snapshots of the z-component of the electric field in the 2D-simulations.
\end{itemize}

\begin{acknowledgement}

This work was supported by the European Research Council (ERC-2024-SyG-101167294 ; UnMySt), the Cluster of Excellence Advanced Imaging of Matter (AIM), Grupos Consolidados (IT1249-19) and SFB925. We acknowledge support from the Max Planck-New York City Center for Non-Equilibrium Quantum Phenomena. The Flatiron Institute is a division of the Simons Foundation. C. M. Bustamante thanks the Alexander von Humboldt-Stiftung for the financial support from the Humboldt Research Fellowship. M.S. acknowledges support by the Office of Naval Research, Grant No. N000142512090. F.P.B. acknowledges financial support from the European Union’s Horizon 2020 research and innovation program under the Marie Sklodowska-Curie Grant Agreement no. 895747 (NanoLightQD).

\end{acknowledgement}

\bibliography{bibliography}

\end{document}